# Electronic structure of the heavy-fermion superconductor $Ce_2Ni_3Ge_5$ and its reference $Ce_2Ni_3Si_5$ compound by *ab initio* calculations

M.J. Winiarski, M. Samsel-Czekała*

*Institute of Low Temperature and Structure Research, Polish Academy of Sciences, P.O. Box 1410, 50-950 Wrocław 2, Poland*

**Abstract**

Band structures of the pressure-induced, heavy-fermion superconductor $Ce_2Ni_3Ge_5$ and its non-superconducting, mixed-valence isostructural (*Ibam*) counterpart $Ce_2Ni_3Si_5$ have been calculated employing the full-potential local-orbital code. Both the local density approximation (LDA) and LDA+$U$ approaches were applied. These investigations were focused particularly on the topology of the Fermi surfaces (FSs) of the compounds. The results show that the FSs are quite similar in these systems and exist in four bands, containing three-dimensional holelike and electronlike sheets. However, the specific FS nesting properties has been revealed only in $Ce_2Ni_3Ge_5$. They support a previously postulated presence of antiferromagnetic spin fluctuations (SF) in the heavy-fermion superconducting state of this germanide under pressure. Such SF can be responsible for the pressure-induced unconventional superconductivity in this system.

## 1. Introduction

The orthorhombic germanide $Ce_2Ni_3Ge_5$ exhibits unusual magnetic properties [1,2,3] leading to the pressure-induced superconductivity around the quantum critical point (QCP) where simultaneously an antiferromagnetic (AFM) order vanishes and a heavy-fermion (HF) state is formed [3,4,5]. An unconventional superconductivity (SC) in HF compounds can be mediated by spin (or magnetic) fluctuations (SF). It is known that quantum SF may be present around the QCP. Interestingly, quite a recent detection of BCS-like SC in $YNiGe_3$ [6], being a non-magnetic analogue to the unconventional pressure-induced superconductor $CeNiGe_3$ [7], have also supported the idea of SF mediated SC in the heavy-fermion Ce-Ni-Ge ternaries [4].

So far only two cerium and nickel ternaries possessing the *Ibam* crystal structure have been reported, these are investigated here $Ce_2Ni_3Si_5$ [8] and $Ce_2Ni_3Ge_5$ [9]. The former silicide is a valence-fluctuator [10,11,12] and a paramagnetic [13] member of the $R_2Ni_3Si_5$ family (where $R$ = rare-earth atom). Most of these rare-earth nickel silicides exhibit an AFM ordering [14,15]. Only $Y_2Ni_3Si_5$ [10] and a BCS-like superconductor $Lu_2Ni_3Si_5$ [16] are paramagnets down to the lowest temperatures measured. The electronic structures of these ternaries were investigated by us earlier [17]. The non-superconducting $Ce_2Ni_3Si_5$, possessing an open *4f*-electron shell, has been chosen in this work as a better analogue to the considered here $Ce_2Ni_3Ge_5$ superconductor than the above $(Y;Lu)_2Ni_3Si_5$ systems.

Previous reports indicated that the AFM structure of $Ce_2Ni_3Ge_5$ with the magnetic unit cell (u.c.) doubled along the *b*-axis direction, has either a collinear (along the *a*-axis [18]) or canted (in the *ab*-plane [19]) arrangements of the local *4f* moments on Ce atoms. The HF state with QCP in $Ce_2Ni_3Ge_5$ is induced by the critical pressure of 3.9 GPa [4] where SC does occur below $T_{SC}$ = 0.26 K. Other superconductors existing in the Ce-Ni-Ge system as $CeNiGe_3$ [7] and $CeNi_2Ge_2$ [20,21] have slightly higher critical temperatures: $T_{SC}$ = 0.48 and 0.43 K, respectively.

In this paper, electronic structures of the $Ce_2Ni_3Ge_5$ and $Ce_2Ni_3Si_5$ systems are reported for the first time. We present densities of states (DOS) and band plots as well as the topology of the Fermi surfaces (FSs) of both investigated compounds in their non-magnetic states. The main aim of this work is a comparison of these two systems, focusing mainly on a possible FS nesting that could be responsible for the SF mechanism of the pressure-induced SC in $Ce_2Ni_3Ge_5$.

## 2. Computational methods

Band structure calculations for $Ce_2Ni_3Ge_5$ and $Ce_2Ni_3Si_5$ ternaries, crystallizing in the orthorhombic *Ibam* structure, displayed in Fig. 1, have been performed with full-potential local-orbital (FPLO-9) method [22]. The Perdew-Wang form of local density approximation (LDA) of the exchange-correlation functional [23] was employed in the scalar and fully [24] relativistic modes. Strongly correlated Ce *4f* electrons were also treated within the LDA+*U* approach [25] (available only in the scalar relativistic mode of FPLO-9), with a typical value of *U* = 6 eV, being applied in other Ce-Ni-Ge [26] and Ce-based [27] ternaries. For $Ce_2Ni_3Ge_5$, powder neutron diffraction data of the lattice parameters: *a* = 0.98085, *b* = 1.18378, and *c* = 0.59602 nm and atomic positions in the u.c. were taken from Ref. [18]. In turn, for $Ce_2Ni_3Si_5$ x-ray diffraction values of lattice parameters: *a* = 0.95651, *b* = 1.11284, and *c* = 0.56453 nm were used [10]. Because of the lack of experimental atomic positions for $Ce_2Ni_3Si_5$, they were assumed as those in other rare-earth nickel silicide ($Y_2Ni_3Si_5$ [8]). The discussion about the selection of the atomic positions in the 235-type ternary nickel silicides has been carried out elsewhere [17]. Here the u.c. is equivalent to double formula units (f.u.). For both studied compounds, the valence-basis sets were selected automatically by the FPLO-9 internal procedure. Total energy values were converged with accuracy to ~1 meV for the 16x16x16 (621 points) *k*-point mesh in the the Brillouin zone (BZ) (or its non-equivalent part).

## 3. Results and discussion

Our calculated scalar relativistic DOS plots within the LDA approach for the $Ce_2Ni_3(Ge;Si)_5$ compounds are presented in Fig. 2. It is seen in this figure that in both systems the main contribution to the DOS at the Fermi level ($E_F$), $N(E_F)$, originates from the Ce *4f* states. Their peaks are nearly identical and almost completely localized above $E_F$ with only small-intensity tails at $E_F$. At the same time, the Ni *3d* and Ge *4p* (or Si *3p*) states have significantly lower values of $N(E_F)$. In both investigated ternaries, the overall shape of the DOSs below $E_F$ is dominated by the Ni *3d* electrons, similarly to their behavior in other nickel intermetallics, e.g. $Lu_2Ni_3Si_5$ and $Y_2Ni_3Si_5$ [17]. However, the centers of masses of these contributions for the Ge- and Si-based systems, being located at -2.0 and -2.5 eV, respectively, are shifted from each other by about 0.5 eV. Also the bottoms of the valence bands of $Ce_2Ni_3Si_5$ are placed slightly deeper below $E_F$ than those in $Ce_2Ni_3Ge_5$. This is mainly due to the lower in energy contribution of the Si *3p* electrons with regards to the Ge *4p* ones.

The differences between scalar and fully relativistic LDA as well as scalar relativistic LDA+*U* results of DOSs in the $Ce_2Ni_3Ge_5$ superconductor are presented in Fig. 3. Interestingly, as seen in this figure, the spin-orbit splitting of the Ce *4f* states of 0.5 eV affects only these states lying above the Fermi level. Hence, the DOSs at $E_F$ (and also Fermi surface) remain unchanged. On the contrary, the calculated within the LDA+*U* value of DOS at $E_F$ (5.0 electrons/eV/f.u.) is slightly lower than the LDA results (6.8 electrons/eV/f.u.). This effect can be explained by the fact that the partial DOSs of the Ce *4f* states are significantly broadened and shifted by the Coulomb repulsion potential *U* to higher energies above $E_F$.

The main influence of an application of the the potential *U* on the LDA electronic structure of $Ce_2Ni_3Ge_5$ is shown in Fig. 4. The conduction bands around the Fermi level are downshifted by about merely 0.01-0.05 eV, which slightly changes the shape of some Fermi surface sheets (presented below). Although assumed here a large value of *U* = 6 eV is consistent with other studies on Ce-based intermetallics, e.g. [26,27], the correct value of the *U* potential for the investigated here superconductor remains still unknown. Bearing in mind that such a large value of *U* causes only minor changes around $E_F$, it is clear that the pure LDA approach may also be sufficient for investigations of electronic structures of other Ce-Ni-Ge compounds, e.g. $Ce_3Ni_2Ge_7$, $CeNiGe_3$ [28], $CeNi_2Ge_2$ [29]. Quite a significant hybridization between the Ce *4f* and *d/p* electron orbitals of Ni and Ge atoms can explain the situation that the correct value of the *U* potential for the Ce *4f* states may be even critically low (<0.5 eV) as that deduced for $CeRu_2Al_{10}$ due to its verification by the valence-band x-ray photoemision spectrum [30]. Nevertheless, since this work is particularly focused on very subtle properties of the FS, both results (LDA and LDA+*U*) are discussed below.

The overall Fermi surface of $Ce_2Ni_3Ge_5$, displayed in Fig. 5 a), is almost the same (in the scale of this figure) as that in $Ce_2Ni_3Si_5$ (not shown). It consists of four three-dimensional FS sheets: two holelike (I, II) and two electronlike (III, IV) ones. It seems that small holelike pockets occurring around the X point in FS sheet I as well as an electronlike pocket around the Γ point in FS sheet IV are irrelevant because of their relatively small volumes. In contrast, large and quite complex FS sheets II and III ought to be responsible for the physical properties of the investigated here intermetallics.

A comparison between the FS sections of $Ce_2Ni_3Ge_5$ and $Ce_2Ni_3Si_5$ through the basal plane, being presented in the figure parts 5 b) and c), reveals pronounced but imperfect nesting properties of FS sheets II and III, possessing almost parallel parts spanned by approximately the same vector **q**. It should be underlined here that these features are exhibited only by the considered here superconductor. Interestingly, the nesting vector **q**=(0.0, 0.5, 0.0)x(2π/*b*) is consistent with the propagation vector of the AFM structure in $Ce_2Ni_3Ge_5$. The imperfect nesting with this vector **q** yields a possibility of appearing, in the non-magnetic HF state around the QCP, spin fluctuations having an antiferromagnetic character rather than the ordered AFM state which might be stabilized only by the perfect nesting. The SF may be tuned by pressure as suggested earlier by Nakashima et al. [4]. These electronic-structure features turn out to be crucial for arising HF superconductivity in $Ce_2Ni_3Ge_5$, especially because the reference Si-based cerium compound apparently does not exhibit any nesting properties of that kind. It should be mentioned here that the SF of an AFM type driven by the nesting properties have also been considered as crucial for SC phenomenon in iron-based superconductors (see e.g. [31,32]) and cuprates (see e.g. [33]). Thus, a pairing mechanism of HF and other unconventional high $T_C$ superconductors is suspected to be similar.

As is visible in Fig. 5 b), the influence of the *U* potential, applied to the Ce *4f* states in the $Ce_2Ni_3Ge_5$

superconductor, on the FS sections (LDA+$U$ results) is meaningless, as it was deduced from the DOS and band plots (Figs. 3 and 4). Note from those figures that the energy broadening of the Ce *4f* states takes place mainly above $E_F$. Therefore, the imperfect nesting properties of II and III FS sheets in this system are quite stable, despite a slight modification of the FS sections. Furthermore, also the overall FS topology in $Ce_2Ni_3Ge_5$ obtained within the LDA+$U$ approach remains also almost unchanged compared with the LDA results. Consequently, the above comparison between the electronic structures of both Ce-based intermetallics, concentrated mainly on the LDA data, seems to be justified.

## 4. Conclusions

The differences found between the electronic structures of $Ce_2Ni_3Ge_5$ and $Ce_2Ni_3Si_5$ can explain a variety in the physical properties of both systems. The pressure-induced heavy-fermion superconductivity in the considered here Ge-based compound may be enabled by spin fluctuations that are driven by imperfect nesting properties of its Fermi surface. These conditions are not present in the reference non-superconducting Si-based compound. Our findings may encourage to further studies on an electronic structure of other heavy-fermion superconductors and reveal some connection between nesting properties of the Fermi surface and a strength of spin fluctuations, tuned by pressure.

**Acknowledgments**

The National Center for Science in Poland is acknowledged for financial support of Project No. N N202 239540. Calculations were carried out in Wroclaw Center for Networking and Supercomputing (Project No. 158). The Computing Center at the Institute of Low Temperature and Structure Research PAS in Wrocław is also acknowledged for the use of the supercomputers and technical support.

**References**

[1] Z. Hossain, S. Hamashima, K. Umeo, T. Takabatake, C. Geibel, F. Steglich, Phys. Rev. B 62 (2000) 8950 .

[2] A. P. Pikul, D. Kaczorowski, P. Rogl, Physica B 312/313 (2002) 422; A. P. Pikul, D. Kaczorowski, P. Rogl, Yu. Grin, phys. stat. sol. (b) 236 (2003) 364.

[3] A. Thamizhavel, H. Nakashima, Y. Obiraki, M. Nakashima, T.D. Matsuda, Y. Haga, K. Sugijama, T. Takeuchi, R. Settai, M. Hagiwara, K. Kindo, Y. Ōnuki, J. Phys. Soc. Jpn. 74 (2005) 2843.

[4] M. Nakashima, H. Kohara, A. Thamizhavel, T. D. Matsuda, Y. Haga, M. Hedo, Y. Uwatoko, R. Settai, Y. Ōnuki, J. Phys.: Condens. Matter 17 (2005) 4539.

[5] M. Nakashima, H. Kohara, A. Thamizhavel, T.D. Matsuda, Y. Haga, M. Hedo, Y. Uwatoko, R. Settai, Y. Ōnuki, Physica B 378-380 (2006) 402.

[6] A.P. Pikul, D. Gnida, Solid State Commun. 151 (2011) 778.

[7] M. Nakashima, K. Tabata, A. Thamizhavel, T.C. Kobayashi, M. Hedo, Y. Uwatoko, K. Shimizu, R. Settai, Y. Ōnuki, J. Phys.: Condens. Matter 16 (2004) L255.

[8] B. Chabot, E. Parthé, J. Less-Common Met. 97 (1984) 285.

[9] A. V. Morozkin, Yu. D. Seropegin, J. Alloys Compd. 237 (1996) 124.

[10] C. Mazumdar, R. Nagarajan, S.K. Dhar, L.C. Gupta, R. Vijayaraghavan, B.D. Padalia, Phys. Rev. B 46 (1992) 9009.

[11] Ch. Mazumdar, R. Nagarajan, C. Godart, L. C. Gupta, B. D. Padalia, R. Vijayaraghavan, J. Appl. Phys. 79 (1996) 6347.

[12] Ch. Mazumdar, Z. Hu, G. Kaindl, Physica B 259 (1999) 89.

[13] R.J. Cava, A.P. Ramirez, H. Takagi, J.J. Krajewski, W.F. Peck, J. Magn. Magn. Matter. 128 (1993) 124.

[14] C. Mazumdar, A.K. Nigam, R. Nagarajan, L.C. Gupta, C. Godart, B.D. Padalia, G. Chandra, R. Vijayaraghavan, Phys. Rev. B 54 (1996) 6069.

[15] C. Mazumdar, K. Ghosh, R. Nagarajan, S. Ramakrishnan, B. D. Padalia, L.C. Gupta, Phys. Rev. B 59 (1999) 4215.

[16] C. Mazumdar, K. Ghosh, S. Ramakrishnan, R. Nagarajan, L.C. Gupta, G. Chandra, B. D. Padalia, R. Vijayaraghavan, Phys. Rev. B 50 (1994) 13879.

[17] M. Samsel-Czekała, M.J. Winiarski, Intermetallics 20 (2012) 63.

[18] L. Durivault, F. Bouree, B. Chevalier, G. Andre, J. Etourneau, J. Magn. Magn. Mater. 246 (2002) 366.

[19] F. Honda, N. Metoki, T.D. Matsuda, Y. Haga, A. Thamizhavel, Y. Okuda, R. Settai, Y. Ōnuki, J. Alloys Compd. 451 (2008) 504.

[20] F.M. Grosche, S.J.S. Lister, F.V Carter, S.S Saxena, R.K.W. Haselwimmer, N.D Mathur, G.G. Lonzarich, Physica B 239 (1997) 62.

[21] F.M. Grosche, P. Agarwal, S.R. Julian, N.J. Wilson, R.K.W. Haselwimmer, S.J.S. Lister, N.D Mathur, F.V Carter, S.S Saxena, G.G. Lonzarich, J. Phys.: Condens. Matter 12 (2000) L533.

[22] FPLO9.00-34, improved version of the FPLO code by K. Koepernik and H. Eschrig, Phys. Rev. B 59 (1999) 1743.

[23] J.P. Perdew and Y. Wang, Phys. Rev. B 45 (1992) 13244.

[24] H. Eschrig, M. Richter, I. Opahle, Relativistic Solid State calculations. In: P. Schwerdtfeger, editor. Relativistic electronic structure theory, part 2. Applications. Theoretical and computational chemistry, vol. 14. Elsevier; 2004. p. 723-76.

[25] H. Eschrig, K. Koepernik, I. Chaplygin, J. Solid State Chem. 176 (2003) 482.

[26] C. Fuente, A. Moral, D.T. Adroja, A. Fraile, J.I. Arnaudas, J. Magn. Mang. Mater. 322 (2010) 1059.

[27] J. Goraus, A. Ślebarski, M. Fijalkowski, phys. stat. Sol. (b) 248 (2011) 2857.

[28] S.F. Matar, B. Chevalier, O. Isnard, J. Etourneau, J. Mater. Chem. 13 (2003) 916.

[29] D. Ehm, F. Reinert, G. Nicolay, S. Schmidt, S. Hufner, R. Claessen, V. Eyert, C. Geibel, Phys. Rev. B 64 (2001) 235104.

[30] J. Goraus, A. Ślebarski, J. Phys.: Condens. Matter 24 (2012) 095503.

[31] W. Bao, Y. Qiu, Q. Huang, M. A. Green, P. Zajdel, M.R. Fitzsimmons, M. Zhernenkov, S. Chang, M. Fang, B. Qian, E.K. Vehstedt, J. Yang, H.M. Pham, L. Spinu, Z.Q. Mao, Phys. Rev. Lett. 102 (2009) 247001.

[32] K. Terashima, Y. Sekiba, J.H. Bowen, K. Nakayama, T. Kawahara, T. Sato, P. Richard, Y.-M. Xu, L. J. Li, G. H. Cao, Z.-A. Xu, H. Ding, T. Takahashi, Proc. Nat. Sci. Acad. 106 (2009) 7330.

[33] D.S. Dessau, Z.-X. Shen, D.M. King, D.S. Marshall, L.W. Lombardo, P.H. Dickinson, A.G. Loeser, J. DiCarlo, C.-H. Park, A. Kapitulnik, W.E. Spicer, Phys. Rev. Lett. 71 (1993) 2781.

**Figures:**

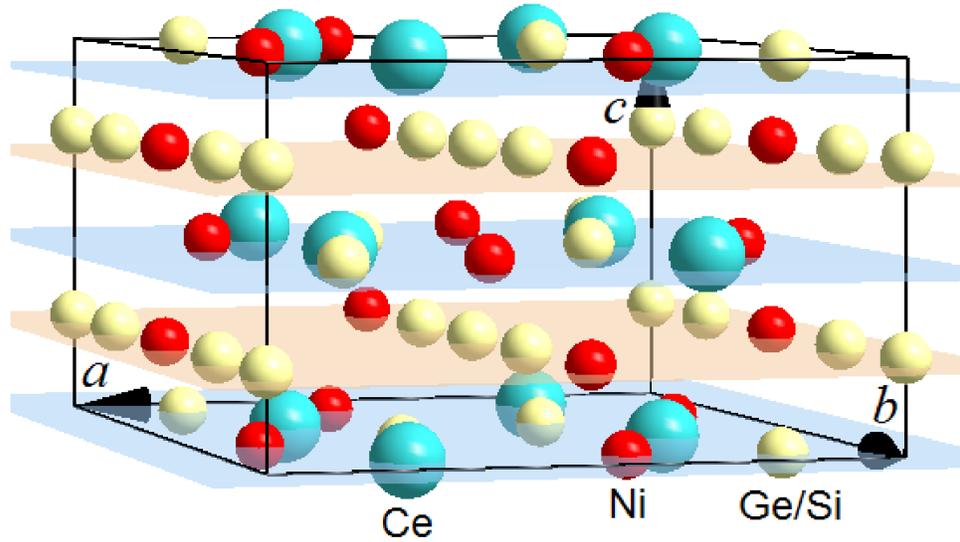

**Fig. 1.** Orthorhombic unit cell of $Ce_2Ni_3(Ge;Si)_5$ systems of the $U_2Co_3Si_5$-type (*Ibam*, space group no. 72).

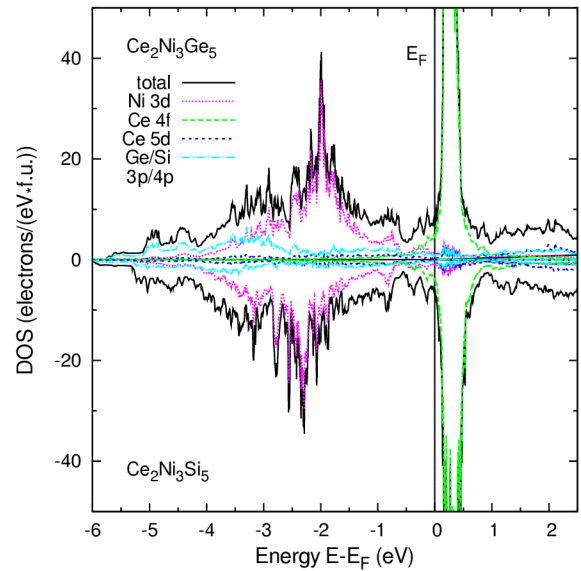

**Fig. 2.** Calculated (scalar relativistic LDA) total and partial (per electron orbitals) DOSs for $Ce_2Ni_3Ge_5$ and $Ce_2Ni_3Si_5$, displayed in upper and bottom parts, respectively.

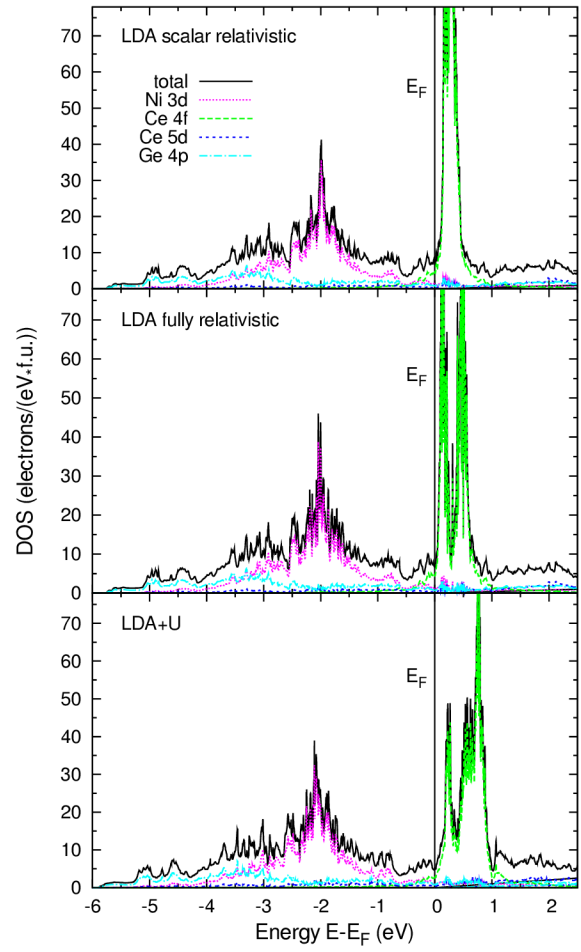

**Fig. 3.** Same as in Fig. 2 but for $Ce_2Ni_3Ge_5$, obtained within scalar or fully relativistic LDA and scalar relativistic LDA+$U$ (with $U$ = 6 eV) approaches.

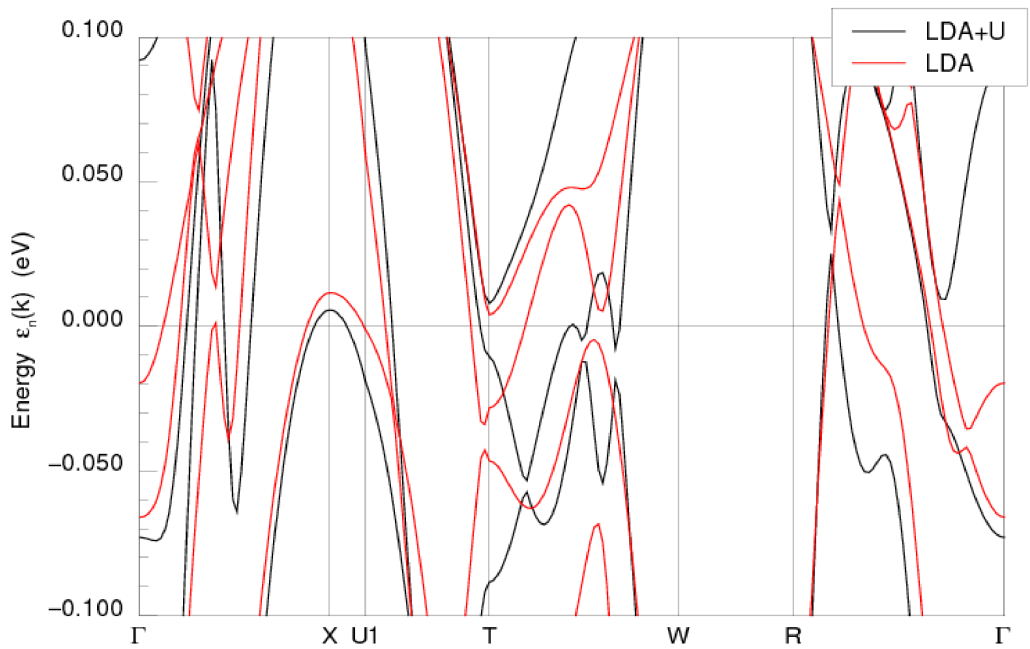

**Fig. 4.** Computed (scalar relativistic LDA and LDA+$U$) band structures of $Ce_2Ni_3Ge_5$ near the Fermi level.

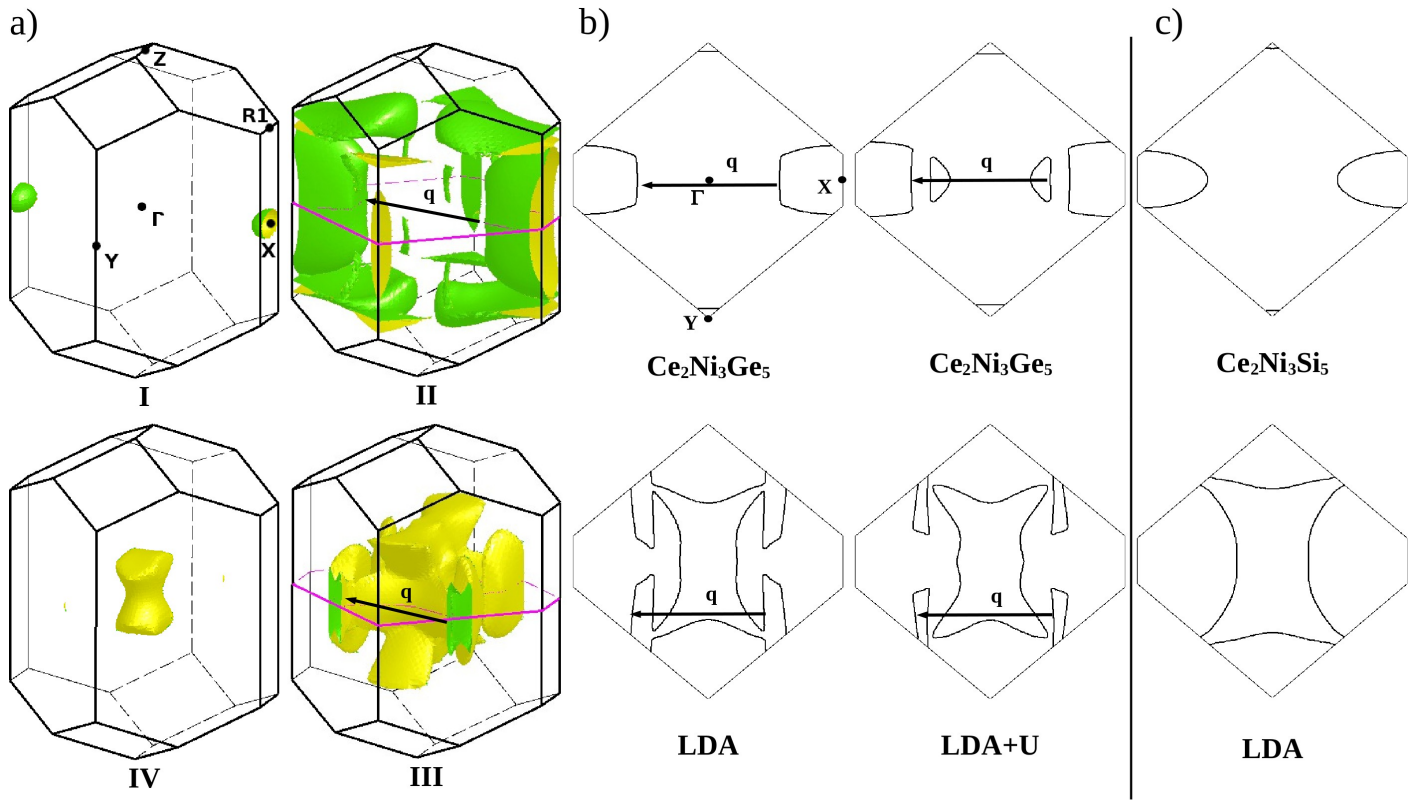

**Fig. 5.**

a) Calculated Fermi surface (LDA) of $Ce_2Ni_3Ge_5$, containing holelike (I and II) and electronlike (III and IV) sheets;

b) Sections of FS sheets II and III in $Ce_2Ni_3Ge_5$ through the basal plane for both LDA and LDA+$U$ results with marked (by arrow) possible nesting vector **q**=(0.0, 0.5, 0.0)x($2\pi/b$).

c) Same FS sections as in b) but for $Ce_2Ni_3Si_5$ in LDA approach. Note the lack of any nesting properties.